\documentclass[aps,prb,twocolumn,showpacs,floatfix]{revtex4-1}

\usepackage{graphicx}
\usepackage{amsmath}

\newcommand{\elautor}[1]{#1 {\it et al.}}

\newcommand{\dep}[2]{\frac{\partial #1}{\partial #2}}
\newcommand{\uta}{Universit\'e de Toulouse; INSA-CNRS-UPS, LPCNO,
 135, Av. de Rangueil, 31077 Toulouse, France}
\newcommand{\utb}{Universit\'e de Toulouse; UPS-CNRS-INSA, IMT,
 118, route de Narbonne, 31062 Toulouse cedex 04, France}
\newcommand{\uam}{Departamento de Ciencias B\'asicas,
 Universidad Aut\'onoma  Metropolitana-Azcapotzalco,
 Av. San Pablo 180, Col. Reynosa Tamaulipas, M\'exico D.F., M\'exico}
\newcommand{\cnrs}{CNRS-LPN, Route de Nozay, 91460 Marcoussis, France}
\begin{document}

\title{Giant spin-dependent photo-conductivity in GaAsN dilute nitride
 semiconductor}

\author{A. Kunold$^{1}$, A. Balocchi$^{2}$, F. Zhao$^{2}$, T. Amand$^{2}$,
        N. Ben Abdallah$^{3}$, J. C. Harmand$^{4}$, X. Marie$^{2}$}
\affiliation{$^1$\uam\\ $^2$\uta \\ $^3$\utb \\ $^4$\cnrs}

\begin{abstract}
A theoretical and experimental study of the
spin-dependent photoconductivity in dilute Nitride GaAsN
is presented.
The non linear transport model we develop here is based
on the rate equations for electrons, holes, deep paramagnetic
and non paramagnetic centers both under CW and pulsed optical excitation.
Emphasis is given to the effect of the competition between paramagnetic
centers and non paramagnetic centers which allows us to reproduce the
measured characteristics of the spin-dependent recombination power dependence.
Particular attention is paid to the role of an external magnetic
field in Voigt geometry. The photoconductivity exhibits a Hanle-type
curve whereas the spin polarization
of electrons shows
two superimposed Lorentzian curves with different widths, respectively related
to the recombination of free and trapped electrons.
The model is capable of reproducing qualitatively and quantitatively
the most important features of
photoluminescence and photocurrent experiments and is helpful
in providing insight on the various mechanisms involved in the electron
spin polarization and filtering in GaAsN semiconductors.
\end{abstract}

\maketitle

\section{Introduction}

The dependence of the recombination time on the relative spin orientation
of photogenerated carriers on paramagnetic
centers, namely the spin dependent recombination (SDR)
has been known for over 50 years since the first
optically detected magnetic resonance (ODMR) experiments carried out by
\elautor{Geschwind}
\cite{geschwind:545} in Al$_2$O$_3$.
It has been observed in surface centers
on crystalline silicon\cite{lepine:805,lepine:436} and on
several systems including
dislocated silicon \cite{wosinski:K57},
amorphous silicon \cite{solomon:505} and
later on in  (Al)GaAs \cite{weisbuch:141,paget:931,weyers:l853}.
In dilute nitride  GaAsN samples SDR has been evidenced with record high values at room temperature
in photoluminescence (PL) experiments \cite{egorov:13539,kalevich:455,lombez:252115,kalevich:174,kalevich:208}
and recently in photoconductivity (PC) measurements\cite{zhao:241104},
stimulating new applications of the SDR as a light- or electron spin-polarization detector.
Furthermore it has been shown by ODMR
that it is in fact Nitrogen-induced Gallium self interstitial deep paramagnetic centers,
and not Nitrogen itself, which harness the SDR in GaAsN alloys\cite{wang:198}.

The key point in the SDR effect is the existence of deep centers inside the gap possessing an
unpaired electron at equilibrium (paramagnetic centers). As a consequence, if a
photogenerated electron in the conduction band and the resident electron on the deep center
have the same spin, the photogenerated electron cannot be captured by the center
(it is generally assumed that the triplet levels are not bound). On the contrary,
when the photogenerated electrons and the the deep center resident electron have
antiparallel spins, the capture is efficient. Therefore the recombination time
of photocreated electrons depends on the relative orientation of the free electron and
of the unpaired electron resident on the center.
After the capture the centers  annihilate one of the two electrons through
a spin-independent recombination with an unpolarized hole.
Under the influence of
linearly polarized light an equal number of spin up and spin down conduction band electrons
are created which are
likewise captured by the paramagnetic centers.
However, under circularly polarized light one given conduction band electron spin orientation
is preferred over the other. This not only creates spin polarized conduction band electrons
through the usual selection rules
but also dynamically polarizes centers. By both removing conduction band electrons
with opposite spin orientation and by blocking the equally polarized ones, this mechanism gives
rise to very high and very long living conduction band electron spin polarization and
to larger PL intensities under circularly polarized light
compared to those under linearly polarized
excitation\cite{weisbuch:141,paget:931,kalevich:455,lombez:252115}.

The theoretical work on SDR can be divided into two groups: stochastic Liouville
equation\cite{kaplan:51,haberkorn:505}, mostly devoted to silicon,
and rate equations\cite{paget:931,weisbuch:141,kalevich:455,
lagarde:208,wang:198,ivchenko:1008.4755v1}
which have been used to study SDR in silicon as well as in GaAs(N).
The former focuses mainly in the process of formation
and dissociation of electron pairs in centers while the latter studies
the conduction band carrier recombination into centers and valence band states.
\elautor{Haberkorn}\cite{haberkorn:505} studied the process of formation
and dissociation of paired spin-up and spin-down electrons via the solution of the
stochastic Liouville equation including the Zeeman
part of the Hamiltonian and a stochastic linear term. Similar models also include
interaction terms between the electron spins. These models are suitable to the
understanding of ODMR and similar experiments.
\elautor{Weisbuch}\cite{weisbuch:141} and
\elautor{Kalevich}
\cite{kalevich:455,kalevich:174,kalevich:208}
studied the recombination dynamics of carriers through a rate equations model that
considers spin up and spin down electrons
coupled to a distribution of paramagnetic centers constituting the SDR channel.
In later approaches an equivalent set of rate differential
equations for the spin variables of electrons and centers as well as the
populations of free and trapped carriers was proposed\cite{kalevich:174,kalevich:208},
lacking however the recombination through non paramagnetic centers which is
important to interpret the power dependence of SDR phenomena.

In this paper we present a comprehensive theoretical and
experimental work on SDR related phenomena in GaAsN, mainly focusing
on photocurrent experiments.
We have developed a new non-linear transport model based on rate equations
for electrons, holes and deep paramagnetic centers in GaAsN
based on the initial model proposed
to explain the strong SDR effects measured in
photoluminescence experiments \cite{kalevich:174,kalevich:208,wang:198,lagarde:208,zhao:174211}.
We show that the electron recombination to non-paramagnetic centers, which was
neglected in the first models, is essential to understand the SDR dependence
as a function of the excitation power.
The proposed theoretical model is able to reproduce
accurately the most important features of both PC and PL experiments.
Finally we have measured and modeled the variation of the spin-dependent photoconductivity and
photoluminescence signals as a function of an external magnetic field in Voigt configuration.
As expected, the SDR effect decreases when the amplitude of the external field
increases. The spin-dependent photoconductivity signal exhibits clearly a Hanle like
curve, well described by the theoretical model.

The paper is organized as follows.
The next section presents the samples' characteristics and the experimental setup.
We begin by discussing a previously formulated rate equations model\cite{kalevich:174,kalevich:208}
in Sec. \ref{model} which is suitable to understand PL experiments.
Starting from this system of differential equations we develop
the model to include non SDR channels in the form of non paramagnetic centers
in Sec. \ref{nonsdrchann} and a uniform external magnetic field in Sec. \ref{bfield}.
These preliminary parts put the fundamentals for the interpretation and modelization
of the SDR mediated transport phenomena developed in Sec. \ref{transport}.
Here we put forward a nonlinear system of partial differential equations by
introducing drift terms for the
conduction band electrons and valence band holes.
Sec. \ref{results} begins with a brief description of the numerical methods
used to solve the system of partial differential equations obtained in the
previous section. We discuss here two main results on transport experiments
and calculations: the SDR as a function of
the incident laser power and applied magnetic field.
Our main results are summarized in Sec. \ref{conclusions}.

\section{Samples and Experimental set-up}\label{exper}

The samples under study were grown under the same conditions
by molecular beam epitaxy at
$T=410^{\circ}{\rm C}$ on
(001) semi-insulating GaAs substrates, after a $400$nm
GaAs buffer layer. They consist of GaAsN semiconductor
epilayers of different Nitrogen content and thickness.
Sample A has a nominally undoped $50$nm thick
GaAs$_{0.981}$N$_{0.019}$ epilayer,
sample B has a nominally undoped $100$nm thick GaAs$_{0.9924}$N$_{0.0076}$
epilayer and sample C a $50$nm thick Si modulation doped (doping density $10^{18}$ cm$^{-3}$)
GaAs$_{0.979}$N$_{0.021}$ epilayer.
The growth was terminated with a $10$nm GaAs cap layer and no post
growth rapid thermal annealing was performed. We have observed similar effects in other doped
or undoped samples with N composition varying in the
range 0.7\% to 2.6\%.

The excitation light was provided by a Ti:Sa laser, propagating along the
direction normal to the sample surface, either in mode-locked regime,
yielding the generation of $1.5$ps pulses at a repetition frequency of
$80$MHz, or in continuous wave operation, at a wavelength $\lambda_{exc}=840$nm.
The laser was focused to a $\sim 150\mu$m or $\sim 40\mu$m diameter spots
respectively in between two Ag electrodes deposited onto the sample surface
$\sim 0.8$mm apart or between two ohmic contacts of variable shape and dimension
defined by lithographic techniques and contacted by ultrasonic wire bonding. The laser light
used for the photo excitation is either circularly polarized (right $\sigma^+$ or left $\sigma^-$)
or linearly ($\sigma^X$ or $\sigma^Y$) polarized along a direction perpendicular
to the surface of the sample ($\boldsymbol{k}$) and modulated by a mechanical chopper.
The sample conductivity was then measured using a synchronous lock-in amplifier from
the voltage drop at the terminals of a $10 k\Omega$ load resistor placed in series with
the sample.
A constant voltage in the range $0<V<12$ volts was applied between the sample electrodes.
The PL intensity was simultaneously measured with the same lock-in technique by recording
its total intensity filtered of the laser scattered light, and  integrated by
a silicon p-i-n photodiode. All the experiments have been performed at room temperature.

\section{Model}\label{model}

In GaAsN samples illuminated by linearly polarized light,
identical populations of spin-up and spin-down electrons are
photogenerated in the conduction band and are subsequently captured
by paramagnetic centers containing an electron with opposite spin. The newly formed
pairs dissociate at the centers living the same population of spin up and spin down
paramagnetic centers as well.
On the contrary, under circularly polarized
excitation $\sigma^-$ ($\sigma^+$) approximately 75\% of the photogenerated electrons
are spin-up (spin-down) polarized and 25 \% are spin-down (spin-up) polarized given
the optical selection rules
in GaAs\cite{meier:1984}. Thus, most of the processes
occurring in the SDR channel will correspond, under $\sigma^+$ excitation,
to spin down conduction electrons rapidly
recombining in spin up paramagnetic traps living most of the spin-down
centers unpaired. After a few cycles the population of
spin-up traps will be considerably depleted. This has the effect of a
double spin filter: first, photogenerated spin-down electrons are prevented from recombining
and second, electrons which undergo spin flipping mechanism are rapidly removed from the
conduction band by the traps owing to the high availability of matching centers.
Moreover, the minority spin-up centers can efficiently trap a spin-down electron increasing the centers' spin polarization, so that 
the whole spin system is self-containing.
At this stage the SDR channel is almost closed and the electrons-hole recombination through band to band
optical transitions  becomes more probable, increasing the
photoluminescence yield. Moreover, as a result of the highly populated conduction band,
a higher photoconductivity is expected under circular polarization compared
to linear excitation.

We start with the rate equations model originally proposed
by 
\elautor{Kalevich}\cite{kalevich:174,kalevich:208,lagarde:208} which
grasps the essential features of the SDR effect:
\begin{align}
\dep{n_\pm}{t}&=\frac{n_\mp-n_\pm}{2\tau_s}
  -\gamma_en_\pm N_\mp +G_\pm,\label{orimodel:eq1}\\
\dep{p}{t}&=-\gamma_hp N_2+G,\label{orimodel:eq2}\\
\dep{N_{\pm}}{t}&=\frac{N_{\mp}-N_{\pm}}{2\tau_{sc}}
  -\gamma_en_\mp N_\pm+\frac{\gamma_hpN_{2}}{2},\\
\dep{N_{2}}{t}&=\gamma_e\left(n_+N_-+n_-N_+\right)-\gamma_hpN_{2}\label{orimodel:eq6}.
\end{align}
Here the subindindices $+$ and $-$ stand for spin polarized
electrons with spin projection $+\hbar/2$ and $-\hbar/2$ respectively
along the incident light axis, thus $n_{+}$ and
$n_{-}$ are the number of spin-up and spin-down conduction band electrons
of total number $n=n_{+}+n_{-}$. Similarly $N_{+}$ and $N_{-}$
are the the spin-up and spin-down paramagnetic traps.
The total number of unpaired paramagnetic traps is given by
$N_1=N_++N_-$ and $N_2$ is the total number of electrons singlets
hosted by the paramagnetic traps.
Holes ($p$) are considered unpolarized as their spin relaxation
time is shorter than $1$ps at room temperature\cite{hilton:146601}
due to the fourfold degeneracy of the valence band at $k = 0$
and the large spin-orbit coupling in III-V semiconductors.
Eqs. (\ref{orimodel:eq1})-(\ref{orimodel:eq6}) ensure conservation of charge
neutrality and number of centers:
\begin{eqnarray}
n-p+N_2=0\label{chgcons:eq1},\\
N_1+N_2=N\label{cencons:eq1}.
\end{eqnarray}
The terms of the form $-\gamma_en_\pm N_\mp$ are responsible for the
spin dependent free electron capture in paramagnetic centers with recombination
rate $\gamma_e$ while the terms $-\gamma_hpN_2$ model the
spin independent recombination of one electron of the paramagnetic center singlet with a hole.
The structure of these terms guarantees that spin-up conduction band electrons
will only be trapped by spin-down centers and viceversa in order to
comply with the Pauli exclusion principle.
This recombination is at the heart of SDR mechanism: under $\sigma^+$ polarized
excitation, for instance,
the $N_-$ population builds up
and the $N_+$ population is symmetrically quenched due to the spin dependent
electron recombination on paramagnetic centers.
Values of the spin relaxation time of free and unpaired electrons on the centers
$\tau_{s}=130ps$ and $\tau_{sc}=2000ps$ respectively and
the effective hole life time $\tau_{h}=35ps$
as well as the
typical ratio of the electron to the hole recombination coefficients
$\gamma_{e}/\gamma_{h}=10$ ( where $\gamma_{h}=1/\tau_{h}N$) are estimated from
previous time resolved PL experiments \cite{lagarde:208,kalevich:455}.

The photo generation of electrons is accounted for by
the terms $G_{+}$ and $G_{-}$ and of holes by $G=G_++G_-$.
Under CW excitation
the generation term per unit surface  is taken as
\begin{equation}\label{gentermcw}
G_\pm\left(t\right)=\frac{\alpha L P}{2\hbar\omega_\nu}
\left\{1 \pm
S_0\left[1+\tanh\left(\frac{t-t_0}{s_t}\right)\right]\right\}
\end{equation}
where $P$ is the laser irradiance, $h\nu$ is the photon's energy,
$2S_0$ is the maximum polarization degree of
the generated electrons, $\alpha$ is the absorption coefficient
of GaAs and $L$ is the width of the GaAsN layer.
The excitation light polarization is gradually switched from linear ($S_0=0$) to circular ($S_0=1/2$)
by the $\tanh$ function; the $s_t$ parameter specifies the switching
time between the the linearly ($S_z$=0) and circularly polarized excitation ($S_z=S_0$).
given the optical selection rules for GaAs under a non resonant excitation involving only
the valence band $s=3/2$ levels, the polarization of the
generation term is set to $S_0=1/4$ when the polarization of light is $100\%$
circular.
Under pulsed excitation the generation terms
are given by
\begin{equation}\label{gentermpulse}
G_\pm\left(t\right)=\frac{\alpha L P t_{pp}}{\sqrt{2\pi}\hbar\omega_\nu s_t}
\left(1\pm 2S_0\right)
{\rm e}^{-\frac{\left(t-t_0\right)^2}{2s_t^2}},
\end{equation}
where $P$ is the laser's time average irradiance
for pulses of period $t_{pp}$ and width $s_t$.
In our experiments $t_{pp}=12$ns and $s_t=1.5$ps.

In CW conditions the total photoluminescence intensity under linear ($X$)
or circular ($+$) excitation is calculated as
$I_{X,+}=\gamma_r n\left(t\right)p\left(t\right)$ where $\gamma_r$ is the optical
recombination rate and $t$ is a sufficiently
long time to insure that the steady state has been reached.
Under pulsed excitation the photoluminescence intensity is calculated
as the time average
$I_{X,+}=\gamma_r\int_0^{t_{pp}} n\left(t\right)p\left(t\right)dt/t_{pp}$. The total
PL acts only as a probe of SDR effect here, since it weakly modifies the $n$ and
$p$ densities\cite{note3}.

A convenient pointer of SDR related phenomena is the SDR ratio ($\mathrm{SDR}_r$) which is
given by the quotient of the total photoluminescence intensity under circularly polarized excitation
to the total photoluminescence intensity under linearly polarized excitation.
The photoluminescence SDR ratio is thus defined as
\begin{equation}\label{sdrrpl}
\mathrm{SDR}_r=\frac{I_+}{I_X}.
\end{equation}
In the absence of the SDR mechanism,
$\mathrm{SDR}_r=1$ whereas in its presence $\mathrm{SDR}_r>1$. In photocurrent experiments
we replace the photoluminescence intensity by the photocurrent in the definition of
$\mathrm{SDR}_r$ given that larger photocurrents are expected in the presence of circularly polarized
light.

\begin{figure}
\includegraphics[width=8.00 cm]{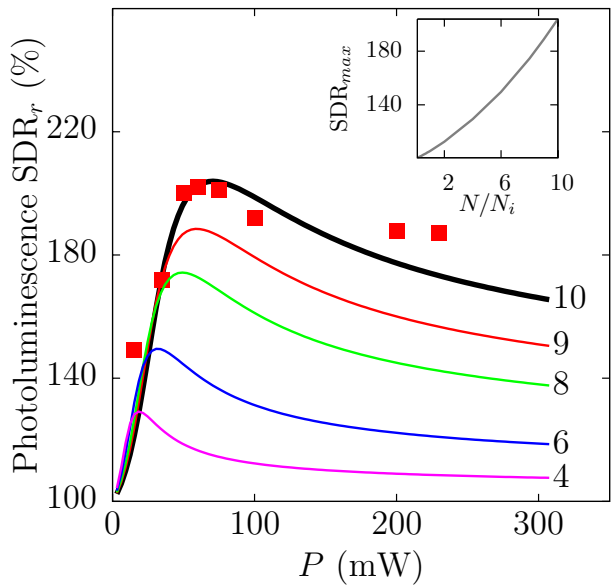}
\caption{(color on line) Photoluminescence SDR ratio vs laser power $P$ for
various ratios of the paramagnetic to non paramagnetic center of
$N/N_i=4$ (purple), $6$ (blue), $8$ (green), $9$ (red) and $10$ (black).
The (red) squares indicate the experimental
results. The dependence of the SDR maximum
SDR$_{max}$ as a function of $N/N_i$ is shown in the inset.}
\label{figurec}
\end{figure}

\subsection{Non-SDR channels: Non paramagnetic centers}\label{nonsdrchann}
Although the main features of the
SDR effect in dilute Nitrides such as the very long conduction electrons
effective spin life time (in time resolved photoluminescence experiments)
\cite{lagarde:208,wang:198}
are well described by the model above, it fails to
reproduce the power dependence of the SDR ratio\cite{zhao:174211}.
Fig. \ref{figurec} shows the photoluminescence measurement [(red) squares] for sample A
in the pulsed excitation regime
and theoretical calculations [(black) solid line] of the $\mathrm{SDR}_r$
as a function of the laser power. The data shows a steep increase of the
$\mathrm{SDR}_r$ as a function of the incident power, up to an optimum value. Above this
power, the $\mathrm{SDR}_r$ smoothly but monotonously decreases.
This is clearly
due to the competition between SDR and non-SDR channels.
In the low power regime ($P<50$mW)
the SDR mechanism dynamically polarizes the spin of paramagnetic centers thus blocking the
recombination of spin polarized conduction band electrons with the same spin orientation
and filtering out oppositely spin polarized ones.
Once the optimum power is reached
($P\approx 75$mW in Fig. \ref{figurec}) most of the paramagnetic centers are
spin polarized. For higher laser powers the SDR mechanism is overwhelmed by
the larger number of generated conduction band electrons and the
recombination through non paramagnetic centers dominates.

This competition mechanism
is a non linear contribution that has an important role
at high excitation power and was not considered in the
first models developed to interpret the remarkable spin
properties of GaAsN\cite{kalevich:174,kalevich:208,lagarde:208}.
The presence of such centers is attested by the observation of numerous
peaks in infrared spectroscopy \cite{wang:241904}.
Introducing this mechanism into the previous set of rate equations we obtain
\begin{align}
\frac{\partial n_\pm}{\partial t}
  &=\frac{n_\mp-n_\pm}{2\tau_s}
  -\gamma_en_\pm N_\mp\nonumber \\
  &-\gamma_{a}n_\pm\left(N_i-N_3\right)
  +G_\pm,\label{model:eq1}\\
\dep{p}{t}&=-\gamma_hp N_2
  -\gamma_{b}pN_3
  +G,\\
\dep{N_{\pm}}{t}&=\frac{N_{\mp}-N_{\pm}}{2\tau_{sc}}
               -\gamma_en_\mp N_\pm+\frac{\gamma_hpN_2}{2},\\
\dep{N_2}{t}&=\gamma_e\left(n_+N_-+n_-N_+\right)-\gamma_hpN_2,\\
\dep{N_3}{t}&=\gamma_a\left(N_i-N_3\right)n
-\gamma_bpN_3
\label{model:eq6}.
\end{align}
Here $N_i$ and $N_3$ are the number of total and occupied non-paramagnetic centers
respectively;
$\gamma_a$ is the recombination rate of conduction band electrons to non
paramagnetic centers and $\gamma_b$ is the recombination rate of nonparamagnetic
centers to valence band.

Assuming that the non paramagnetic
centers have the same recombination rates
as paramagnetic centers (i.e. $\gamma_a/\gamma_e\approx 1$
and $\gamma_b/\gamma_h\approx 1$), the main adjustable parameter of our model is
the total number of
non paramagnetic centers.
Their role is clearly evidenced in Fig. \ref{figurec}. In this
graph we present a series of $\mathrm{SDR}_r$ vs power curves for various rates
of paramagnetic centers to non paramagnetic centers
$N/N_i$ keeping $N$ constant. As $N/N_i$ increases, the $\mathrm{SDR}_r$ maximum increases and
moves to higher irradiance values.
As the number of non paramagnetic centers is increased, the effective
conduction electron
recombination time $1/\gamma_a N_i$ [see Eq. (\ref{model:eq1})]
is shortened making this mechanism more likely at lower laser irradiance. At the
same time the SDR maximum ratio is reduced as a consequence of the increasing
transit of non polarized electrons through the non paramagnetic centers \cite{wang:198}.
The best agreement
with the experimental results for sample A (Fig. \ref{figurec})
is achieved with $N_i=N/10$
where the total paramagnetic center density $N=N_+ + N_- + N_2$ is
kept constant. The fitting of the PL $\mathrm{SDR}_r$ power dependence allows us to fix
the ratio $N/N_i$ to be used in the following photoconductivity model.

\subsection{Effect of a magnetic field}\label{bfield}
In order to include magnetic field effects into the model, we first
recall the simple case of a spin 1/2 particle coupled to the
magnetic field via the Zeeman Hamiltonian
\begin{equation}\label{zeeman:eq1}
H=\hbar \boldsymbol{\omega} \cdot {\boldsymbol{s}}
\end{equation}
where $\boldsymbol{\omega}= g \mu_B \hbar\boldsymbol{B}$, $g$ is the
gyromagnetic factor of free electrons,
$\mu_B$ is the Bohr magneton and ${\boldsymbol{s}}$ is the dimensionless
vector spin operator for the particle. By solving the von Neumann equation for
the density matrix and averaging the spin vector
one finds that the Hamiltonian (\ref{zeeman:eq1}) yields the well known
coherent term contribution for the total spin evolution
\begin{equation}\label{precfree:eq1}
\frac{d \boldsymbol S}{dt}=\boldsymbol{\omega}\times\boldsymbol{S}.
\end{equation}
The spin component along the propagation of light direction
is related to
the number of spin-up and spin-down electrons in $\hbar$ units
through $S_z=\left(n_+-n_-\right)/2$\cite{kalevich:174,kalevich:208}.

For the trapped electrons we can work out a similar
dynamical equation given by
\begin{equation}
\frac{d \boldsymbol{S}_c}{dt}=\boldsymbol{\Omega}\times\boldsymbol{S}_c.
\end{equation}
where $\boldsymbol{\Omega}=g_c\mu_B \hbar\boldsymbol{B}$ with $g_c$ the
gyromagnetic factor for trapped electrons and the spin component along the $z$
axis is $S_{zc}=\left(N_+-N_-\right)/2$.
The gyromagnetic factors for free electrons and trapped electrons
were set to $g=0.5$ and $g_c=2$ respectively \cite{kalevich:174,kalevich:208,pettinari:245202,zhao:041911}.

Subtracting Eq. (\ref{model:eq1}) for spin-down to spin-up electrons
and adding the term (\ref{precfree:eq1}) for spin precession
we obtain
\begin{eqnarray}
\frac{d S_z}{dt}+\frac{\gamma_e}{2}\left(n N_1-4 S_zS_{zc}\right)
+\frac{1}{\tau_s}S_z\nonumber\\
+\left(\boldsymbol{S}\times\boldsymbol{\omega}\right)_z=\frac{G_+-G_-}{2}.
\end{eqnarray}
where we have used the identity
\begin{equation}
n_+N_-+n_-N_+=\frac{1}{2}\left(nN_1-4 S_zS_{zc}\right).
\end{equation}

Adding Eq. (\ref{model:eq1}) for spin-up and spin-down electrons
considering isotropic spin relaxation time $\tau_s$ we
obtain the rate equation for the total number of conduction band
electrons. An analogous procedure is followed for the paramagnetic
center variables.

Finally the whole set of rate equations for the spin variables,
total conduction band and trapped electrons is given by
\begin{align}
\dot{\boldsymbol{S}}
  &+\frac{\gamma_e}{2}\left(\boldsymbol{S}N_1-\boldsymbol{S}_cn\right)
  +\frac{\gamma_a}{2}\boldsymbol{S}\left(N_i-N_3\right)\nonumber\\
  &+\frac{1}{\tau_s}\boldsymbol{S}+\boldsymbol{S}\times\boldsymbol{\omega}
  =\boldsymbol{\Delta G},\label{mag:eq1}\\
\dot{\boldsymbol{S}}_c
  &+\frac{\gamma_e}{2}\left(\boldsymbol{S_c}n-\boldsymbol{S}N_1\right)
  +\frac{1}{\tau_{sc}}\boldsymbol{S_c}
  +\boldsymbol{S}\times\boldsymbol{\Omega}
  =\boldsymbol{0},\\
\dot{n}
  &+\frac{\gamma_e}{2}\left(nN_1-4\boldsymbol{S}\cdot\boldsymbol{S}_c\right)
  \nonumber \\
&+\gamma_a n\left(N_i-N_3\right)=G,\\
\dot{p}&+\gamma_hN_2p-\gamma_bpN_3=G,\\
\dot{N_1}
  &+\frac{\gamma_e}{2}\left(nN_1-4\boldsymbol{S}\cdot\boldsymbol{S}_c\right)
  -\gamma_hN_2p=0,\\
\dot{N_2}
  &-\frac{\gamma_e}{2}\left(nN_1-4\boldsymbol{S}\cdot\boldsymbol{S}_c\right)
  +\gamma_hN_2p=0,\\
\dot{N_3}&=\gamma_a\left(N_i-N_3\right)n
-\gamma_bpN_3
\label{mag:eq6}
\end{align}
where $\boldsymbol{\Delta G}=(G_+-G_-)\boldsymbol{k}/2$.
In an isotropic media the dynamical equations for the three spin
components should be invariant under rotations, hence we have
replaced $nN_1-4 S_zS_{zc}$ by $nN_1-4 \boldsymbol{S}\cdot\boldsymbol{S}_{c}$.
By doing this
the recombination rate of conduction electrons to paramagnetic centers
is increased when $\boldsymbol{S}$ and $\boldsymbol{S}_c$ are antiparallel whereas
it vanishes when they are parallel as expected from the Pauli exclusion
principle needed to form a singlet state.

\subsection{Photoconductivity}\label{transport}
Having defined the most important generation terms we turn
to the central problem of this work:  electron transport combined with SDR effects.

Modulation of the photoconductivity with the polarization of the excitation
light up to $40\%$ has been indeed measured very recently at
$T=300$ K \cite{zhao:241104}.

Several theoretical contributions have been presented
to study the transport of electrons in doped GaAsN\cite{fahy:035203}.
\elautor{Fahy} used a tight binding model
and the Boltzmann equation in the relaxation-time approximation
to obtain the mobility
of conduction band electrons in GaAsN\cite{fahy:035203}. Nevertheless
the spin degree of freedom was not considered in this work.

In order to focus on SDR associated phenomena we present a
simpler approach in which the electron and hole density currents
are introduced through drift terms for the current neglecting
other transport phenomena as diffusion.
The Einstein-Smoluchowski relation for the diffusion
coefficient yields a diffusion length
of a few hundred nanometers compared
to the laser spot size in the order of the $100 \mu m$
due to the low mobility ($\mu\approx 100$ cm$^2$V/s\,\,\,\cite{fahy:035203,li:113308,mouillet:333})
of electrons and holes in GaAsN. Therefore diffusion currents are neglected
in the calculation.

In this case $n_\pm$ and $p$ are understood as the
1D local densities of electrons and holes,
therefore $n_\pm\equiv n_\pm\left(t,x\right)$ and $p\equiv p\left(t,x\right)$.
Similarly the number of paramagnetic impurities can be replaced by
the local density of impurities $N_\pm\equiv N_\pm\left(t,x\right)$
and $N_2\equiv N_2\left(t,x\right)$.
$\boldsymbol{S}\equiv \boldsymbol{S}\left(t,x\right)$ and
$\boldsymbol{S}_c\equiv \boldsymbol{S}_c\left(t,x\right)$
are the free electron and paramagnetic trap spin densities
respectively.

The drift density current for electrons is
$J_n=J_++J_-$ where $J_{\pm}=-\mu_eE n_\pm$ corresponds
to spin up- and spin down-electrons.
For holes the drift density current is  $J_p=\mu_hE p$.
Here $\mu_e$ and
$\mu_h$ are the electron and hole mobilities respectively.
They are set
to $\mu_{e}=200$ cm$^{2}/Vs$ and $\mu_{h}=100$ cm$^{2}/Vs$
respectively as measured and calculated
for similar samples\cite{li:113308,fahy:035203,mouillet:333,fahy:3731}.
We must also include the spin density current terms.
According to 
\elautor{Dyakonov}\cite{dyakonov:126601},
the spin polarization current density tensor, neglecting a small spin orbit contribution,
can be written for electrons as
\begin{equation}\label{spincurr1}
\boldsymbol{J}_{\boldsymbol{S}}=-\left(\mu_eE +D\frac{\partial}{\partial x}\right)\boldsymbol{S},
\end{equation}
where $D$ is the diffusion coefficient.
Due to the small conduction electron mobility the diffusion term can
be neglected in Eq. (\ref{spincurr1}) so that
\begin{equation}
\boldsymbol{J}_{\boldsymbol{S}}\approx-\mu_eE \boldsymbol{S}.
\end{equation}
Gathering all the terms we obtain the following system of equations
\begin{align}
\dot{\boldsymbol{S}}
  &-\mu_e\left(\frac{\partial E}{\partial x}+E \frac{\partial }{\partial x}\right)\boldsymbol{S}
  +\frac{\gamma_e}{2}\left(\boldsymbol{S}N_1-\boldsymbol{S}_cn\right)
  \nonumber\\
  &+\frac{\gamma_a}{2}\boldsymbol{S}\left(N_i-N_3\right)
   +\frac{1}{\tau_s}\boldsymbol{S}+\boldsymbol{S}\times\boldsymbol{\omega}
  =\boldsymbol{\Delta G},\label{tra:eq1}\\
\dot{\boldsymbol{S}}_c
  &+\frac{\gamma_e}{2}\left(\boldsymbol{S_c}n-\boldsymbol{S}N_1\right)
  +\frac{1}{\tau_{sc}}\boldsymbol{S}_c
  +\boldsymbol{S}\times\boldsymbol{\Omega}
  =\boldsymbol{0}\label{tra:eq2},\\
\dot{n}
  &-\mu_e \left(\frac{\partial }{\partial x}\right)nE
  +\frac{\gamma_e}{2}\left(nN_1-4\boldsymbol{S}\cdot\boldsymbol{S}_c\right)
  \nonumber \\
  &+\gamma_a n\left(N_i-N_3\right)=G,\\
\dot{p}&+\mu_h \left(\frac{\partial }{\partial x}\right)Ep
   +\gamma_hN_2p+\gamma_bpN_3=G,\\
\dot{N_1}
  &+\frac{\gamma_e}{2}\left(nN_1-4\boldsymbol{S}\cdot\boldsymbol{S}_c\right)
  -\gamma_hN_2p=0,\\
\dot{N_2}
  &-\frac{\gamma_e}{2}\left(nN_1-4\boldsymbol{S}\cdot\boldsymbol{S}_c\right)
  +\gamma_hN_2p=0,\\
\dot{N_3}&-\gamma_a\left(N_i-N_3\right)n=0
-\gamma_bpN_3.
\label{tra:eq6}
\end{align}

The density of photogenerated carriers is obtained
by multiplying (\ref{gentermcw}) or (\ref{gentermpulse}) by
the Gaussian spatial distribution
$\exp\left[-\left(x-x_0\right)^2/2s_x^2\right]/\sqrt{2\pi}s_x$
which specify the laser profile needed in transport simulations.
The laser spot radius $s_x$ is varied according to the experimental
conditions.
Even though the generation terms for transport carriers do not figure
explicitly in the equations above, they are considered through the
boundary conditions. The number
of incoming electrons and holes on opposite sides of the sample
are fixed
for a given voltage and electrical current.

It is worthwhile mentioning that, while
the expression for the local conservation of paramagnetic centers
[Eq. (\ref{cencons:eq1})] remains unchanged,
the charge conservation condition [Eq. (\ref{chgcons:eq1})]
is replaced now by the continuity relation
\begin{eqnarray}
\frac{\partial}{\partial t}\left(n-p+N_2+N_3\right)
+\frac{\partial}{\partial x}\left(J_e-J_p\right)=0.
\end{eqnarray}

As for the PL experiments it is convenient to define a
parameter that allows us to pin the SDR phenomena. The transport SDR ratio,
similarly to the photoluminescence SDR ratio, is thus the quotient of the time
averaged density current in circular polarization ($\bar{J}_+$) to the
time averaged density current in linear polarization ($\bar{J}_X$):
\begin{equation}
\mathrm{SDR}_r=\frac{\bar{J}_+}{\bar{J}_X}
\end{equation}
where explicitly $\bar{J}_{+,X}=\int_0^{t_{pp}}J_{+,X}\left(t\right)dt/t_{pp}$.


\begin{figure}
\includegraphics[width=8.00 cm]{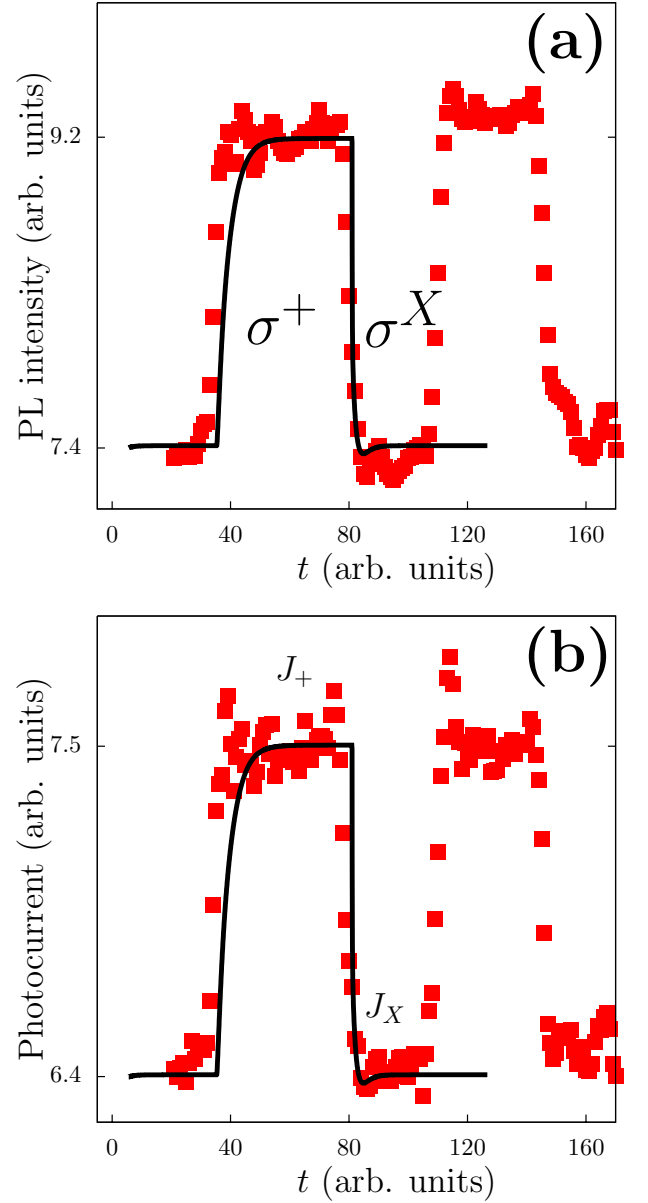}
\caption{(color online) (a) Photoluminescence intensity and (b) photocurrent
signal as a function of time under circularly ($\sigma^+$) and linearly
($\sigma^X$) polarized light.
The (red) squares indicate the experimental results for PL and PC
and the thick black line represents the theoretical calculations.}
\label{figurel}
\end{figure}

\begin{figure}
\includegraphics[width=8.00 cm]{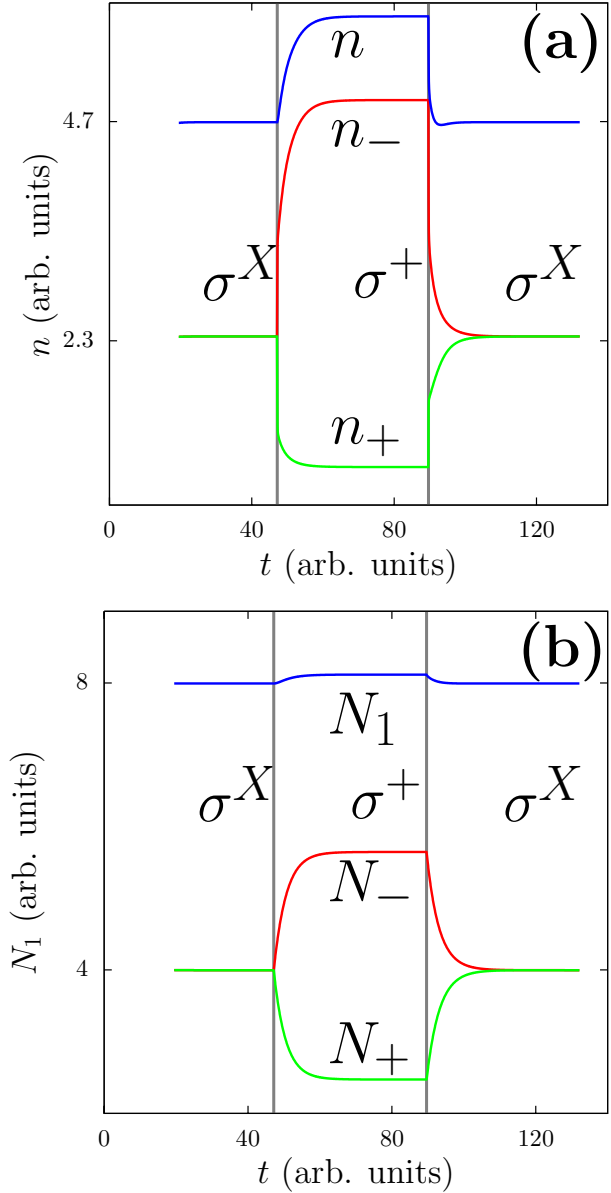}
\caption{(color online)
(a) Spin-up $n_+$ (green), spin-down $n_-$ (red) and total $n$ (blue) density
of free electrons.
(b) Density of trapped electrons on paramagnetic centers:
Spin-up $N_+$ (green), spin-down $N_-$ (red) and $N_1=N_++N_-$ (blue).}
\label{figurell}
\end{figure}

\begin{figure}
\includegraphics[width=8.00 cm]{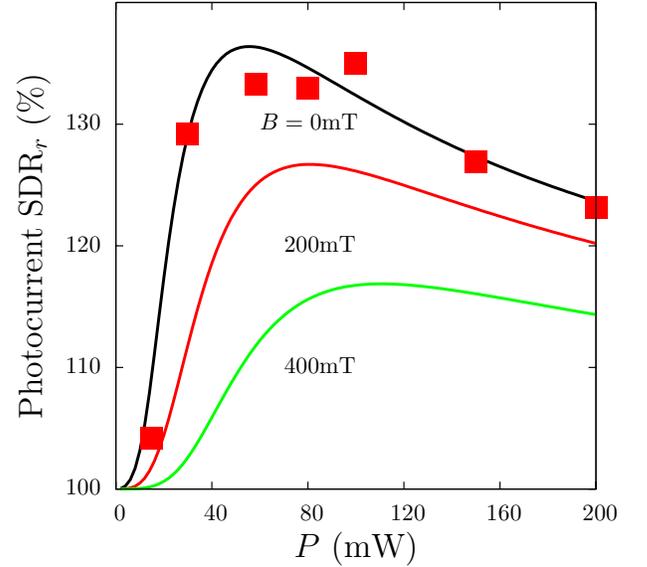}
\caption{(color on line) Photocurrent $\mathrm{SDR}_r$ vs laser irradiance
$P$ for different values of the magnetic field.
The (red) squares are the experimental results, the dark solid lines
indicate the theoretical curves for different transverse magnetic fields
$0$mT (black), $200$mT (red) and $400$mT (green).}
\label{figuref}
\end{figure}

\begin{figure}
\includegraphics[width=8.00 cm]{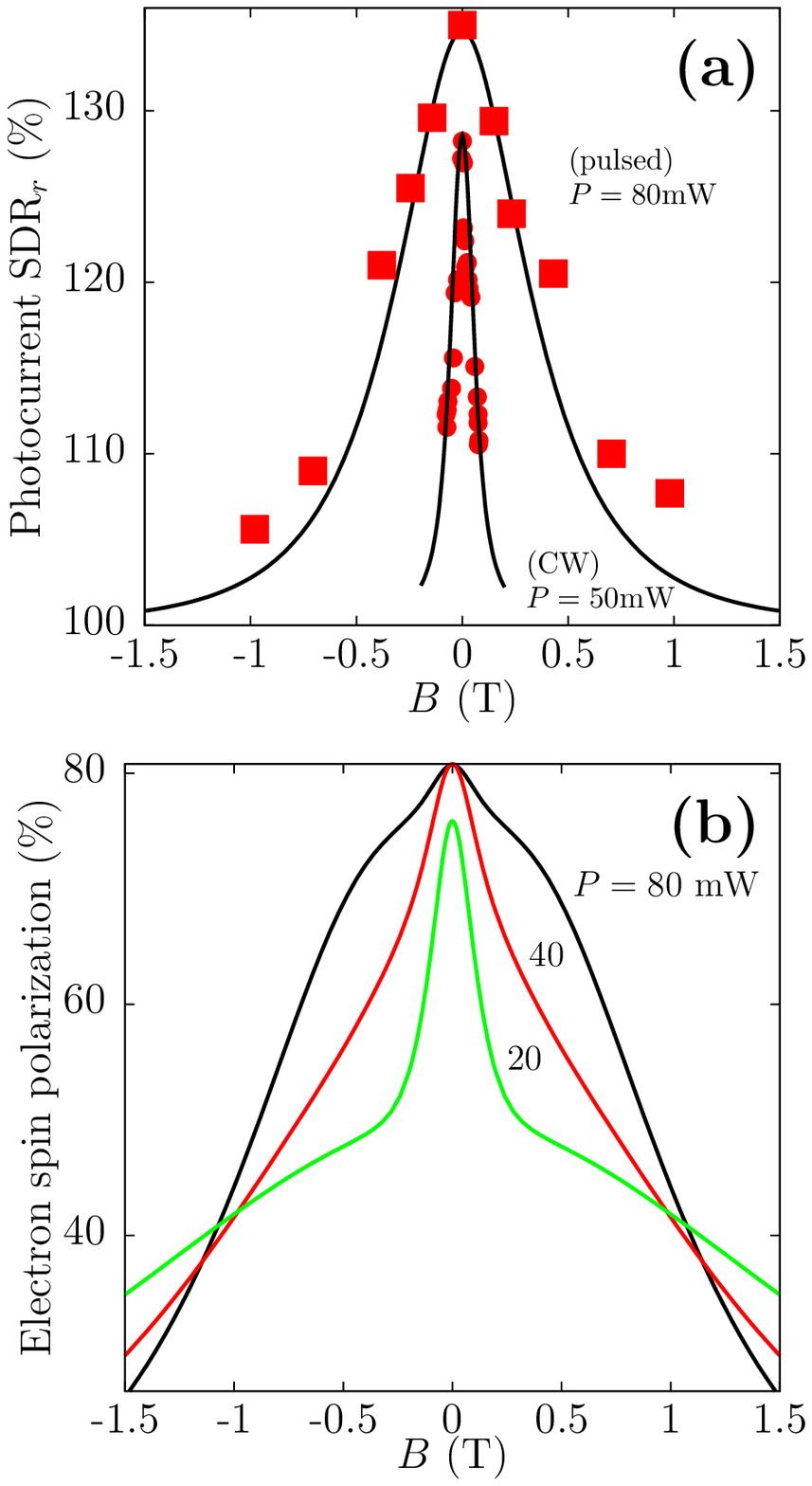}
\caption{(color on line) Hanle curve. (a) Photocurrent SDR ratio vs
magnetic field $B$. The (red) squares display the experimental results
for a power of $80$mW in pulsed excitation and the (red) circles
exhibit the experimental results at $100$mW in CW excitation.
The solid lines correspond to the theoretical calculations.
(b) Theoretical values of the conduction band electron
circular polarization $P_c$
vs magnetic field $B$ for $P=80$ mW (black), $40$ mW (red)
and $20$ mW (green) in pulsed operation.
}
\label{figureg}
\end{figure}

\section{Results and discussion}\label{results}

The rate Eqs. (\ref{mag:eq1})-(\ref{mag:eq6}) were solved by
fourth-order Runge-Kutta method. As initial conditions the amount
of spin-up and spin-down electron on paramagnetic centers was set to $N/2$
while the remaining variables were set to zero.

In the transport case the carrier generation terms
were introduced through the boundary conditions for
a fixed voltage and current intensity;
Dirichlet boundary conditions
were imposed separately at the two ends of the sample for
electrons and holes given that their distributions build up at
opposite extremes.
Accordingly, the space
was discretized by forward and backwards differences for
electrons and holes respectively.
Initially the spin $\boldsymbol{S}_c$ of trapped electrons
was set to zero uniformly along the sample.
As the mobility in GaAsN samples is very low it is possible to neglect
the effects of transport in a first round of calculations using
fourth-order Runge-Kutta method for each discretized position.
The resulting system was solved by point Gauss-Seidel
iteration method including the transport terms. In this calculations
we used the same fitting parameters as the one introduced in Sec. \ref{model}.

In Fig. \ref{figurel} we present the simultaneous measurement
of photoluminescence (a) and photocurrent (b) in sample B with
lithographic contacts under
CW excitation of $P=5$mW.
The (red) squares indicate the experimental measurement of PL and
photocurrent while the solid lines correspond to the
theoretical calculations obtained from the solution of
the set of differential Eqs. (\ref{tra:eq1})-(\ref{tra:eq6}).
In this graph we observe an increase in the PL intensity
under circularly polarized ($\sigma^+$) light compared to
a linearly polarized excitation ($\sigma^X$).
Likewise the photocurrent
increases under a $\sigma^+$ excitation with respect to the
$\sigma^X$ case. A clear correlation is observed
between the PL and PC behaviors.
Under $\sigma^X$ excitation the density of spin-up and spin-down
polarized electrons are identical, namely $n_+=n_-$ 
whereas with a $\sigma^+$ excitation $n_{-}$ becomes considerably larger than
$n_{+}$ [Fig. \ref{figurell} (a)].
Through the dynamical polarization mechanisms the $N_-$ to $N_+$
population ratio correspondingly follows the conduction band electron one [Fig. \ref{figurell} (b)].
When $n_-\gg n_+$, the $N_-$ density sharply increases, blocks further recombination of
conduction band electrons of same spin orientation, and the $N_+$ population is depleted.
Accordingly, for linearly polarized light we can see identical $N_+$ and $N_-$ populations.
The same model reproduces well the dependence of the PL and PC signal as functions
of the excitation light polarization [solid line in Figs. \ref{figurel} (a) and \ref{figurel} (b)].

Fig. \ref{figuref} [(red) solid squares]
shows the photocurrent $\mathrm{SDR}_r$ as
a function of power $P$ under pulsed laser excitation
for sample C whereas the solid (black) line represents
the theoretical photocurrent $\mathrm{SDR}_r$ obtained from the set of differential
Eqs. (\ref{tra:eq1})-(\ref{tra:eq6}).
Features identical to the PL experiments are measured
and are precisely reproduced by the model.

In order to get further insight we studied the
Hanle effect i.e. the
depolarization of electrons in the presence of a magnetic
field $B$ in Voigt configuration\cite{kalevich:4929} and the consequent
quenching of the SDR rate.
For non vanishing values of the magnetic field (red for $200$mT and
green line for $400$mT) a progressive decrease in the $\mathrm{SDR}_r$
is observed in Fig. \ref{figuref} as expected. As the transverse
magnetic field is increased the originally polarized spin of conduction band and trapped electrons
begin to precess around the magnetic field with different angular velocities
given the different $g$ factors. The recombination
through paramagnetic centers is thus triggered due to
the dephasing induced mismatch between spin of conduction band electrons
and trapped electrons.

In Fig. \ref{figureg} (a) we present the experimentally
recorded variation of the SDR ratio as a function of an external
magnetic field for sample C at $P=80$ mW under pulsed excitation conditions.
A very strong reduction of the photocurrent SDR ratio describing
a quasi Lorentzian curve is observed.
The model described in Sec. \ref{bfield}
reproduces accurately the measured magnetic field dependence
[(black) solid line in Fig. \ref{figureg} (a)].
To clarify further the magnetic field dependence let us recall
that
in the simplest situation of an electron
spin precessing around the external
magnetic field, the Hanle effect yields a
Lorentzian dependence for the electron spin polarization described by:
\begin{equation}\label{lorentz}
P_c=\frac{P_0}{1+\left(\frac{B}{B_{1/2}}\right)^2}
\end{equation}
where the half-width
$B_{1/2}$ is given by\cite{meier:1984,kalevich:4929}
\begin{equation}\label{spinreltime}
B_{1/2}=\frac{\hbar}{g\mu_BT_s},
\end{equation}
where $T_s$ is the mean decay time:
\begin{equation}
\frac{1}{T_s}=\frac{1}{\tau_s}+\frac{1}{\tau}
\end{equation}
and $\tau$ is the electron lifetime.
We emphasize here that the data displayed in Fig. \ref{figureg} (a)
do not correspond to a classical Hanle curve which probes the spin
depolarization in a photoluminescence experiment.
Therefore, to interpret the experimental data,
we have calculated in Fig. \ref{figureg} (b) the free
electron spin polarization degree
as a function of the magnetic field $B$ obtained from the
model for various values of the power for circularly
polarized ($\sigma^+$) excitation.
This curve presents new features
hidden in the SDR magnetic field dependence in PC. The curves show two
superimposed Lorentzian with different half widths that
strongly suggest two distinct
processes with different mean spin life times.
The magnetic field has here the double effect of depolarizing the conduction band
and paramagnetic center resident electrons. Due to their significantly different spin
lifetimes the $B$ field dependence of the polarization shows the double Lorentzian curve
as observed by 
\elautor{Kalevich}\cite{kalevich:4929} in CW PL polarization experiments.
For low excitation power the narrow Lorentzian is controlled by the localized electrons (long)
spin life time while the larger Lorentzian corresponds to the free electron (short) spin
life time.
In the photoconductivity experimental results of Fig. \ref{figureg} (a), 
the width of the curve is determined by the localized electron spin lifetime $T_s$
which corresponds to the electron lifetime $\tau$. The latter is controlled by the spin-dependent
capture of free electrons (see Sec. \ref{model}).
To further support this interpretation we have measured the magnetic field dependence of the
photocurrent at much lower photo-generated electron density (i.e. under CW excitation).
As shown in Fig. \ref{figureg} (a) [(red) circles], the width of the curve is one order
of magnitude smaller due to the longer localized electron lifetime\cite{note2}. We emphasize that the same set of parameters were used
in the calculations for Fig. \ref{figurel}, Fig. \ref{figuref} and  Fig. \ref{figureg}.

\section{Conclusions}\label{conclusions}

We have presented photoluminescence and photocurrent measurements in GaAsN samples that
evidence the presence of many SDR related phenomena such as a spin dependent
photoconductivity change and Hanle effect in the SDR ratio of the
photoconductivity.
We have found a strong correlation between photoluminescence and photoconductivity results.
To get further insight on the interplay of the many different mechanisms
involved we developed a model that correctly reproduces the main features
of these phenomena. The model is based on a set of non-linear transport differential
equations containing drift density currents for electrons and holes and terms
that account for photogeneration, SDR in paramagnetic centers,
and recombination through non paramagnetic centers.
The presence of a constant magnetic field is introduced
through spin-precession terms.
First, we have shown that the interplay of SDR and non SDR channels is
responsible for the most important features of the SDR ratio vs laser
irradiance curve. It presents a maximum that can be explained as
a result of the competition between recombination through SDR and non
SDR channels that activate at high laser pumping. As the recombination through non
paramagnetic centers becomes more likely the maximum SDR ratio decreases.
Second, we investigated the effect of a transverse  magnetic
field (Voigt configuration) on the Spin Dependent Photoconductivity. The SDR ratio vs magnetic field
curve shows a Lorentzian-like shape which is  reminiscent
of the Hanle effect experienced by conduction band electrons and traps.

\acknowledgments
This paper is dedicated to the memory of Naoufel Ben Abdallah.
Alejandro Kunold acknowledges financial support from
``Estancias sab\'aticas al extranjero'' CONACyT and ``Acuerdo 02/06'' Rector\'{\i}a UAM-A,
to support his sabbatical stay from UAM-A in INSA Toulouse.
Xavier Marie acknowledges the support of IUF and Cost Action n$^\circ$ MP0805.
The support of
QUATRAIN(BLAN07-2 212988) funded by the French ANR and from the Marie Curie
Project DEASE: MEST-CT-2005-021122 of the European Union is also acknowledged.


\begin{thebibliography}{10}
\bibitem{note3}%
  \BibitemOpen
  \bibinfo {note} {The bimolecular recombination process has thus been
  neglected in Eqs. (\ref{orimodel:eq1})-(\ref{orimodel:eq2}).}%
  \bibAnnoteFile{Stop}{note3}%
\bibitem{note2}%
  \BibitemOpen
  \bibinfo {note} {We measured a similar width of the classical Hanle curve
  determined by the polarization of the CW photoluminescence spectroscopy.}%
  \bibAnnoteFile{Stop}{note2}%
\end{thebibliography}

%

\end{document}